\documentclass[reprint,amsmath,amssymb,superscriptaddress]{revtex4-2}

\usepackage{bm}
\usepackage{graphicx}
\usepackage{amsmath}
\usepackage{amsbsy}
\usepackage{amstext}
\usepackage{amssymb}
\usepackage{units}
\usepackage[utf8]{inputenc}
\usepackage[english]{babel}
\usepackage[colorlinks, citecolor=blue, urlcolor=blue, linkcolor=red,breaklinks]{hyperref}

\begin{document}
	
	\title{Coulomb drag between two graphene layers at different temperatures}
	
	\author{F. Escudero}
	\email{federico.escudero@uns.edu.ar}
	\affiliation{Departamento de Física, Universidad Nacional del Sur, Av. Alem 1253, B8000, Bahía Blanca, Argentina}
	\affiliation{Instituto de Física del Sur, Conicet, Av. Alem 1253, B8000, Bahía Blanca, Argentina}
	\author{F. Arreyes}
	\affiliation{Departamento de Física, Universidad Nacional del Sur, Av. Alem 1253, B8000, Bahía Blanca, Argentina}
	\author{J. S. Ardenghi}
	\affiliation{Departamento de Física, Universidad Nacional del Sur, Av. Alem 1253, B8000, Bahía Blanca, Argentina}
	\affiliation{Instituto de Física del Sur, Conicet, Av. Alem 1253, B8000, Bahía Blanca, Argentina}
	
\begin{abstract}
   We theoretically study the Coulomb drag in graphene when there is a temperature difference between the layers. Within the degenerate limit for equal layer densities, we find that this can lead to significant deviations from the usual quadratic temperature dependence of the drag resistivity. The exact behavior depends strongly on the phase space available for intraband scattering, and is not symmetrical when the temperatures of the layers are interchanged. In particular, when one layer is at a much higher temperature $T$ than the other, the drag resistivity behaves as $\rho_D\sim T/d^5$, where $d$ is the interlayer separation. The magnitude of the drag in this limit is always larger when the active layer is at the higher temperature.
\end{abstract}

\maketitle  

\section{Introduction}\label{sec:introduction}

Many-body interactions play a crucial role in our understanding of many important phenomena in condensed matter physics. Their effect, however, is usually seen indirectly in conventional transport measurements. One important exception is the Coulomb drag effect between two closely spaced layers \cite{Rojo1999,Narozhny2016}, which directly depends on the interlayer many-body interactions. Experimentally,
the effect consists of driving a current in one layer and measuring
the voltage induced in the other isolated layer. The latter
appears due to the interlayer many-body interactions, with the main mechanism
being the Coulomb interaction. The quantity of interest is usually
the drag resistivity $\rho_{D}=E_{p}/j_{a}$, where $E_{p}$ is the
induced electric field in the passive layer, and $j_{a}$ is the applied
current density in the active layer. 

Originally studied in semiconductor heterostructures and quantum wells \cite{Hoepfel1986,Hoepfel1988,Gramila1991,Solomon1989,Solomon1991},
over the last decade there has been considerable interest in the Coulomb drag between two graphene layers \cite{Tse2007,Kim2011,Kim2012,Narozhny2012,Song2013,Gamucci2014,Li2016}.
As a two-dimensional material, graphene is well known for its relativisticlike dispersion relation at low energies, where electrons behave as chiral massless fermions \cite{CastroNeto2009,Peres2010,DasSarma2011,Craciun2011}. The ability to easily manipulate the carrier density in graphene leads to unique features in the Coulomb drag. This is particularly reflected
at low carrier densities, i.e., close to charge neutrality, where the
drag resistivity follows an unconventional dependence with the chemical
potential $\mu$ and the temperature $T$ \cite{Narozhny2012}. In the degenerate limit $\mu/k_{B}T\gg1$,
both layers are in the Fermi-liquid regime, and the drag resistivity
follows the known quadratic temperature dependence at low temperatures \cite{Hwang2011,Amorim2012,Carrega2012,Lux2012}. 

Theoretical calculations of the Coulomb drag usually assume equal temperature layers. This is motivated by conventional experimental measurements in which the temperature varies uniformly throughout the drag setup, including both layers. Furthermore,
due to the required separation between the layers (usually no
larger than a few hundred nanometers), it seems impractical, at first sight, to consider a temperature difference between them.
However, recent technological
advances at the nanoscale may allow one to create (and maintain)
a temperature difference between the layers \cite{Koebel2012,He2015,Tang2015}. Negligible interlayer
heat transfer may be accomplished using low thermal conductivity materials as a dielectric spacer between the layers \cite{Fuji2015,Gupta2018}. 

In the Fermi-liquid regime, the dependence $\rho_{D}\propto T^{2}$
is usually understood in terms of phase-space considerations:
the momentum transfer between the layers comes from electron-hole
excitations, which are limited by the temperature. When the layers are at different temperatures,
at first sight one may expect $\rho_{D}\propto T_{a}T_{p}$, where $T_a$ and $T_p$ are the temperatures of the active and passive layer, respectively. However, one should keep in mind that in a typical drag experiment the current is driven
through the active layer, where the external electric field is applied. The passive layer is usually isolated, so the drag resistivity only depends on
deviations from equilibrium in the distribution function of the active
layer. This means that the drag effect should, in principle, depend differently on the temperature of each layer. 

In this work, we theoretically investigate the Coulomb drag between two graphene layers at different temperatures.
Within the
Fermi-liquid regime, we find significant deviations from the usual
behavior $\rho_{D}\propto T^{2}$. 
In particular, when the temperature $T_{i}$ of one layer is much
higher than the temperature $T_{j}$ of the other layer,
we find a linear dependence $\rho_{D}\propto T_{i}$, practically independent of $T_{j}$. The magnitude of the effect depends on the situation: for $T_{a}\gg T_{p}$
the drag resistivity is about two times larger than for $T_{p}\gg T_{a}$. This reflects the nonsymmetrical behavior of the drag effect when the temperatures of the layers are interchanged.
Furthermore, the limit $T_{i}\gg T_{j}$ also implies a stronger dependence
on the interlayer separation, $\rho_{D}\propto1/d^{5}$, compared
to the usual dependence $\rho_{D}\propto1/d^{4}$. We show that these
results are strongly related to the allowed intraband transitions
in graphene.

This work is organized as follows: In Sec. \ref{sec:theoretical model} we develop the theoretical
model. The drag conductivity is calculated using the kinetic theory approach. The transport is assumed to be dominated by impurities, with a momentum-dependent scattering time. The nonlinear susceptibilities are calculated within the degenerate limit $\mu/k_BT\gg1$. In Sec. \ref{sec:results} we present the numerical results for the drag resistivity, as a function of the temperature of each layer. We discuss the drag behavior when the temperature of one layer is fixed, and when there is a fixed temperature difference $\Delta T$ between the layers. We also obtain analytical approximations for the drag resistivity in the limit $T_i\gg T_j$. Finally, our conclusions follow in Sec. \ref{sec:conclusions}. 

\section{Theoretical model}\label{sec:theoretical model}

\begin{figure}[t]
	\includegraphics[scale=0.8]{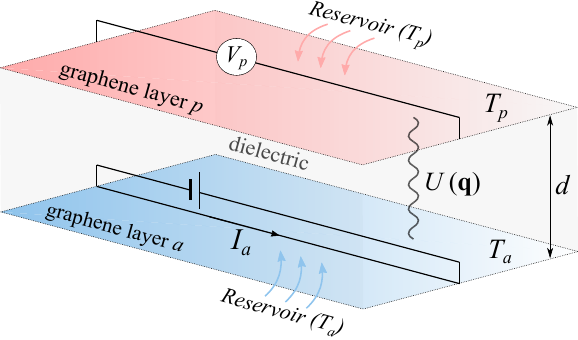}	
	\caption{Schematic configuration of a Coulomb drag setup between two graphene layers. In the active layer $a$ a current $I_a$ is driven, inducing a voltage $V_p$ in the other passive $p$ due to the interlayer Coulomb interaction $U\left(\mathbf{q}\right)$. The drag resistivity is given by $\rho_D\propto V_p/I_a$. The layers are separated by a distance $d$ by a dielectric medium. Each end is in contact with a different thermal reservoir, which induces a temperature difference $T_a-T_p$ between the graphene layers.  }\label{fig:1}
\end{figure}

The system configuration is shown in Fig. \ref{fig:1}. Two graphene layers
are separated by a distance $d$, so that no tunneling between them is
possible. A current $I_{a}$ is passed through the active layer ($i=a$), inducing
a voltage $V_{p}$ in the passive layer ($i=p$). Throughout this work, we consider that both layers are electron doped, and in the Fermi-liquid regime $T_F/T\gg1$. We thus neglect the contribution of plasmons to the drag, which are expected to be relevant at high temperatures $T\gtrsim0.2T_{F}$ \cite{Flensberg1995,Badalyan2012}, and consider the static screening within the random phase approximation \cite{Narozhny2012}.  We
assume, for simplicity, equally doped layers 
and a uniform dielectric
spacer with a relative permittivity $\varepsilon_{r}$ \cite{Katsnelson2011,Carrega2012,Badalyan2012}. The screened interlayer Coulomb interaction is then given by \cite{Kamenev1995}
\begin{equation}
	U_{ap}\left(\mathbf{q}\right)=\frac{e^{2}}{2\varepsilon_{0}\varepsilon_{r}}\frac{q}{\left(q^{2}+2qq_{T}\right)e^{qd}+2q_{T}^{2}\sinh\left(qd\right)},
\end{equation}
where $q_{T}=4\alpha k_{F}$ is the Thomas-Fermi wave vector and
$\alpha\sim2.2/\varepsilon_r$ is the coupling constant in graphene \cite{CastroNeto2009}. 

Since our interest in this work is to keep the layers at different temperatures, we assume that the dielectric medium is a good thermal insulator \cite{Fuji2015}. One possibility is to use aerogel as a nanospacer, which has an extremely low thermal conductivity due to its high porosity and very low density \cite{He2015,Tang2015,Gupta2018}. This may allow one to keep, during the measurements, the graphene layers at different temperatures if each one is in contact with a thermal reservoir (Fig. \ref{fig:1}). 

\subsection{Drag conductivity}\label{subsec:drag conductivity}

To obtain the drag conductivity we will use the kinetic theory approach \cite{Jauho1993,Hwang2011}. The
transport in each layer is assumed to be dominated by impurity
scattering, within the ballistic regime.
In practice this means that the interlayer separation must be much
smaller than the mean free path of the electrons. Consequently, within the relaxation time approximation, to linear order one has \cite{Jauho1993,DasSarma2011}
\begin{equation}
	\tilde{f}_{\mathbf{k},s}^{i}\simeq f_{\mathbf{k},s}^{i}+f_{\mathbf{k},s}^{i}\left(1-f_{\mathbf{k},s}^{i}\right)\psi_{\mathbf{k},s}^{i},\label{eq:noneq}
\end{equation}
where $\psi_{\mathbf{k},s}^{i}=\beta_{i}e\mathbf{E}_{i}\cdot\tau_{\mathbf{k}}^{i}\mathbf{v}_{\mathbf{k},s}$, with $\beta_i=1/k_BT_i$. Here $f_{\mathbf{k},s}^{i}$ and $\tilde{f}_{\mathbf{k},s}^{i}$
are the equilibrium and nonequilibrium distribution functions
in the $i$ layer, $\epsilon_{\mathbf{k},s}=s\hbar v_{F}k$ is the
low energy dispersion relation of electrons in the $s=\pm1$ band, and $\mathbf{v}_{\mathbf{k},s}=\hbar^{-1}\nabla\epsilon_{\mathbf{k},s}$
is the velocity vector. For charged impurities, the momentum-dependent
scattering time reads $\tau_{\mathbf{k}}^{i}=\tau_{i}k$ \cite{CastroNeto2009,DasSarma2011}. For usual
carrier densities in graphene, say $n\sim10^{12}\,\mathrm{cm^{-2}}$, the Fermi-liquid regime holds for temperatures up
to $T\sim100$ K. 

Following standard manipulations \cite{Jauho1993,Flensberg1995,Narozhny2012,Lux2012}, when the layers are at different
temperatures the transconductivity is found to be given by 
\begin{align}
	\sigma_{ij} & =\frac{e^{2}}{16\pi k_{B}T_{j}}\sum_{\mathbf{q}}\left|U_{ap}\left(\mathbf{q}\right)\right|^{2}\nonumber \\
	& \times\int_{-\infty}^{\infty}d\omega\frac{f_{ij}^{\alpha}\left(\mathbf{q},\omega,\Delta\beta\right)}{\sinh\left(\beta_{i}\hbar\omega/2\right)\sinh\left(\beta_{j}\hbar\omega/2\right)},\label{eq:transconductivity}
\end{align}
where $\alpha$ is the current direction, and
\begin{align}
	f_{ij}^{\alpha}\left(\mathbf{q},\omega,\Delta\beta\right) & =\Gamma_{i}^{\alpha}\left(\mathbf{q},\omega\right)\left[e^{\Delta\beta\hbar\omega/2}\Gamma_{j}^{\alpha}\left(\mathbf{q},\omega\right)\right.\nonumber \\
	& \qquad\left.-2\sinh\left(\Delta\beta\hbar\omega/2\right)\tilde{\Gamma}_{j}^{\alpha}\left(\mathbf{q},\omega\right)\right],\label{eq:f}
\end{align}
where $\Delta\beta=\beta_{i}-\beta_{j}$. The nonlinear susceptibilities
read
\begin{align}
	\boldsymbol{\Gamma}_{i}\left(\mathbf{q},\omega\right) & =-2\pi g\sum_{\mathbf{k},s,s'}\left(f_{\mathbf{k},s}^{i}-f_{\mathbf{k}+\mathbf{q},s'}^{i}\right)F_{s,s'}\left(\mathbf{k},\mathbf{q}\right)\nonumber \\
	& \qquad\times\left(\tau_{\mathbf{k}}^{i}\mathbf{v}_{\mathbf{k},s}-\tau_{\mathbf{k}+\mathbf{q}}^{i}\mathbf{v}_{\mathbf{k}+\mathbf{q},s'}\right)\delta_{\mathbf{q},\omega,\mathbf{k}},\label{eq:NLsus0}\\
	\tilde{\boldsymbol{\Gamma}}_{i}\left(\mathbf{q},\omega\right) & =-2\pi g\sum_{\mathbf{k},s,s'}\left(f_{\mathbf{k},s}^{i}-f_{\mathbf{k}+\mathbf{q},s'}^{i}\right)F_{s,s'}\left(\mathbf{k},\mathbf{q}\right)\nonumber \\
	& \times\left[\left(1-f_{\mathbf{k},s}^{i}\right)\tau_{\mathbf{k}}^{i}\mathbf{v}_{\mathbf{k},s}-f_{\mathbf{k}+\mathbf{q},s'}^{i}\tau_{\mathbf{k}+\mathbf{q}}^{i}\mathbf{v}_{\mathbf{k}+\mathbf{q},s'}\right]\delta_{\mathbf{q},\omega,\mathbf{k}},\label{eq:NLsus}
\end{align}
where for brevity we have used the notation $\delta_{\mathbf{q},\omega,\mathbf{k}}\rightarrow\delta\left(\hbar\omega+\epsilon_{\mathbf{k},s}-\epsilon_{\mathbf{k}+\mathbf{q},s'}\right)$.
The factor $g=4$ takes into account the spin
and valley degeneracy. The function 
\begin{equation}
	F_{s,s'}\left(\mathbf{k},\mathbf{q}\right)=\frac{1+ss'\cos\left(\theta_{\mathbf{k+q}}-\theta_{\mathbf{k}}\right)}{2},
\end{equation}
where $\theta_{\mathbf{k}}=\arctan\left(k_y/k_x\right)$,
comes from the pseudo-spinors overlap in graphene \cite{CastroNeto2009,DasSarma2011}. 

Equation (\ref{eq:transconductivity}) is quite general, in the sense that the only approximations
made are the starting assumptions. For equal layer temperatures it reduces to the well-known drag conductivity
in graphene \cite{Tse2007,Kim2011,Narozhny2012}. The function $\boldsymbol{\Gamma}_{i}\left(\mathbf{q},\omega\right)$
is the usual nonlinear susceptibility. The new function
$\tilde{\boldsymbol{\Gamma}}_{i}\left(\mathbf{q},\omega\right)$ only appears in the transconductivity when $T_{i}\neq T_{j}$, and as we shall see, it can lead to
significant deviations from the drag behavior at equal layer temperatures. 

The obtained expression for the transconductivity reflects the two general ways by which the drag depends
on the temperature of the layers \cite{Narozhny2016}. One comes from the available electronic
states for scattering, which results in the known factor
$\sim\sinh^{-1}\left(\beta\hbar\omega/2\right)$ (one for each layer). The other temperature dependence
comes from the deviations from equilibrium in the distribution function
[cf. Eq. (\ref{eq:noneq})]. At equal layer temperatures this simply yields
a prefactor $\sim 1/T_j$ in the transconductivity $\sigma_{ij}$, and the usual
nonlinear susceptibility (\ref{eq:NLsus0}) in both layers (which is temperature independent in
the degenerate limit). When $T_{i}\neq T_{j}$, the response of the
system is modified according to Eq. (\ref{eq:f}). This function reflects the fact that for interlayer scattering between two layers at different temperatures,
the number of states available for scattering into differs from the number of states
available for scattering out, even at equilibrium [see Appendix \ref{appendix:transconductivity}, Eq. (\ref{eq:App1})]. Importantly, the function $f_{ij}^{\alpha}\left(\mathbf{q},\omega,\Delta\beta\neq0\right)$ depends on $T$, even within the degenerate limit $\mu/k_BT\gg1$.

The drag conductivity (\ref{eq:transconductivity}) is not symmetrical when the temperatures of the layers are interchanged. This behavior is expected due to the configuration of the layers in the drag setup: in the active layer a current is driven, whereas the passive layer is isolated. Consequently, the nonequilibrium response is different in each layer. This also implies that
the $2\times2$ conductivity matrix is not symmetrical when $T_{a}\neq T_{p}$, i.e., $\sigma_{ap}\neq\sigma_{pa}$. Note that the asymmetrical behavior of the transconductivity when the layers are interchanged violates the Onsager relation because the two layers are not in thermal equilibrium.

\subsection{Nonlinear susceptibilities}\label{sec:NLsus}

In the Fermi-liquid regime the dominant scattering of electrons
is intraband, which allows us to simplify the calculation of the nonlinear
susceptibilities by considering only the conduction band. Since $f_{ij}^{\alpha}\left(\mathbf{q},\omega,\Delta\beta\right)=f_{ij}^{\alpha}\left(-\mathbf{q},-\omega,\Delta\beta\right)$, it is sufficient to do the calculations for $\omega>0$.
In the limit
$\mu/k_{B}T\gg1$, at low energies and momenta we have \cite{Narozhny2012,Hwang2011} 
\begin{align}
	\boldsymbol{\Gamma}_{i}\left(\mathbf{q},\omega\right) & \simeq\theta\left(v_{F}q-\omega\right)\frac{g\tau_{\mathbf{k}_{F}}^{i}\omega}{\pi\hbar v_{F}}\frac{\mathbf{q}}{q},
\end{align}
where $\theta$ is the step function. To evaluate the other function $\tilde{\boldsymbol{\Gamma}}_{i}$ analytically,
we note that in the limit $\mu/k_{B}T\gg1$ we have $\left(-\partial f_{\mathbf{k}}/\partial k\right)\rightarrow\delta\left(k-k_{F}\right).$
Thus we can take
\begin{align}
	\left(f_{\mathbf{k}}-f_{\mathbf{k}+\mathbf{q}}\right)f_{\mathbf{k}+\mathbf{q}}\sim\frac{k_{F}}{\beta\mu} & f_{\mathbf{k}}\left(1-e^{-\beta\hbar\omega}\right)\delta_{k,k_{F}},\label{eq:aprox1}\\
	\left(f_{\mathbf{k}}-f_{\mathbf{k}+\mathbf{q}}\right)\left(f_{\mathbf{k}}+f_{\mathbf{k}+\mathbf{q}}-1\right) & \sim\frac{k_{F}}{\beta\mu}\left(\delta_{\left|\mathbf{k}+\mathbf{q}\right|,k_{F}}-\delta_{k,k_{F}}\right),\label{eq:aprox2}
\end{align}
where we have used that, by conservation of energy, $\left|\mathbf{k}+\mathbf{q}\right|=k+\omega/v_{F}$.
Then the evaluation of $\tilde{\boldsymbol{\Gamma}}_{i}$ is straightforward;
at low energies and momenta we obtain (cf. Appendix \ref{appendix:NLS})
\begin{align}
	\tilde{\boldsymbol{\Gamma}}_{i}\left(\mathbf{q},\omega\right) & \simeq\theta\left(v_{F}q-\omega\right)\frac{g\tau_{\mathbf{k}_{F}}^{i}\omega}{\pi\hbar v_{F}}\frac{\mathbf{q}}{q}\frac{1}{\beta_{i}\hbar\omega}\nonumber \\
	& \quad\times\left[\tanh\left(\frac{\beta_{i}\hbar\omega}{2}\right)-2\left(\frac{\omega}{v_{F}q}\right)^{2}\right].\label{eq:NLsus2}
\end{align}

\subsection{Drag resistivity}\label{subsec:drag resistivity}

We now focus on the drag resistivity $\rho_{D}$, which is what experiments
usually measure. Inverting the $2\times2$ conductivity matrix yields
\begin{equation}
	\rho_{D}=\frac{\sigma_{pa}}{\sigma_{ap}\sigma_{pa}-\sigma_{aa}\sigma_{pp}}\simeq-\frac{\sigma_{pa}}{\sigma_{aa}\sigma_{pp}},
\end{equation}
where $\sigma_{ii}=e^{2}2\pi v_{F}\tau_{i}n_{i}/h$ is
the single layer conductivity in graphene \cite{CastroNeto2009} (assuming charged impurities).
The last approximation follows from $\sigma_{ij}\ll\sigma_{ii}$,
which implies that in the active layer the drag effects are negligible,
i.e., $j_{a}\simeq\sigma_{aa}E_{a}$. Note that the drag resistivity
depends on the transconductivity $\sigma_{pa}$ (rather than $\sigma_{ap}$) [cf. Eq. (\ref{eq:transconductivity})].

\begin{figure*}[t]
	\includegraphics[scale=0.35]{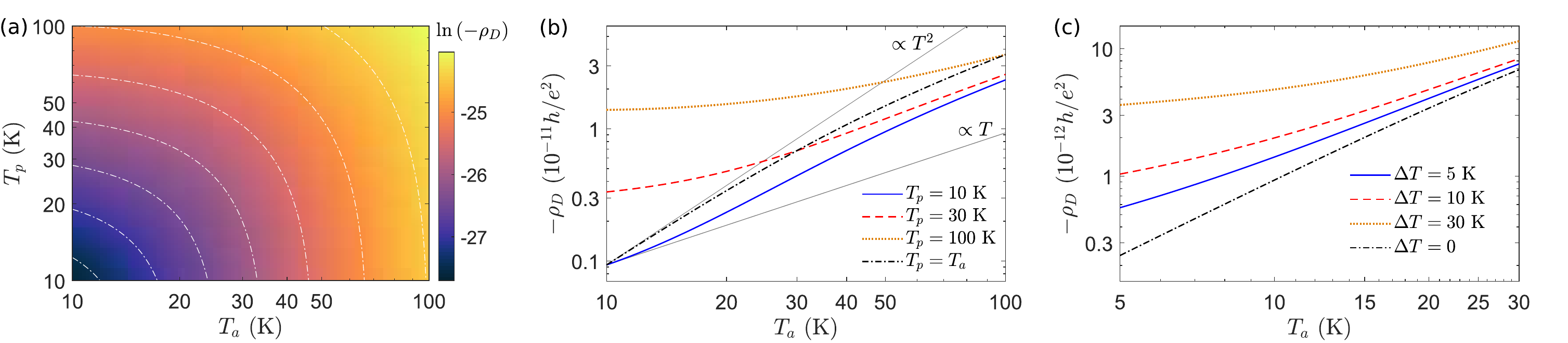}	
	\caption{(a) Density plot of the drag resistivity $\rho_D$ in logarithmic scale (in units of $h/e^2$), as a function of the temperatures $T_a$ and $T_p$ of the active and passive layers, respectively. The white dot-dashed lines correspond to constant values of $\ln\left(-\rho_D\right)$, in steps of 0.5. (b) Drag resistivity as a function of $T_a$, for fixed values of $T_p$. Solid black lines represent $\rho_D\propto T^2$ and $\rho_D\propto T$ behavior. (c) Drag resistivity as a function of $T_a$, for fixed temperature differences $\Delta T=T_p-T_a$. In the last two figures, the black dot-dashed line is the drag resistivity at equal temperatures. In all cases we consider an interlayer separation $d=100$ nm, carrier density $n=10^{12}\,\mathrm{cm^{-2}}$, and coupling constant $\alpha=2$. }\label{fig:2}
\end{figure*}

Introducing adimensional variables $x=dq$ and $y=\hbar\omega\beta_{a}$,
the drag resistivity can be written as
\begin{eqnarray}
	\rho_{D} & = & -\frac{h}{e^{2}}\frac{g^2}{256\pi}\frac{k_{B}T_{a}}{\mu}\frac{k_{B}T_{p}}{\mu}\frac{1}{\left(dq_{T}\right)^{2}}\frac{1}{\left(dk_{F}\right)^{2}}\nonumber \\
	& \times & \int dx\frac{x^{3}e^{-2x}}{\left[\left(x\kappa+1\right)^{2}-e^{-2x}\right]^{2}}G\left(x,\lambda,\gamma\right),\label{eq:resistivity}
\end{eqnarray}
where $\kappa=1/dq_{T}$, and 
\begin{align}
	G\left(x,\lambda,\gamma\right) & =\int_{0}^{x/\lambda}dy\frac{y\gamma}{\sinh\left(y/2\right)\sinh\left(y\gamma/2\right)}\left\{ ye^{-y\left(1-\gamma\right)/2}\right.\nonumber \\
	& +\left.2\sinh\left[\frac{y}{2}\left(1-\gamma\right)\right]\left[\tanh\left(\frac{y}{2}\right)-2\left(\lambda\frac{y}{x}\right)^{2}\right]\right\}.\label{eq:G}
\end{align}
Here $\lambda=dk_{B}T_{a}/\hbar v_{F}$ and $\gamma=T_{a}/T_{p}$.
The upper limit $x/\lambda$ in the integration over $y$ comes from the
allowed intraband transitions ($\omega<v_{F}q$) \cite{CastroNeto2009}, which leads to a linear dependence $\rho_{D}\propto T$ for equal temperature layers and large values of $\lambda$ (i.e., high temperatures and
large interlayer separations) \cite{Lux2012,Chen2015,Escudero2022}. 

Clearly, for arbitrary $T_a$ and $T_p$ the temperature dependence of the drag resistivity (\ref{eq:resistivity}) does not follow a simple power law. The usual Fermi-liquid result $\rho_D\propto T^2$ is obtained only when $\gamma=1$ and $\lambda\ll1$, which implies $G\simeq4\pi^2/3$ \cite{Lux2012}. Note that the prefactor $\sim1/\mu^{2}$ in Eq. (\ref{eq:resistivity}) (for doped layers with equal carrier density;
in general $\sim1/\mu_{a}\mu_{p}$) implies that the drag always requires
electron-hole asymmetry \cite{Narozhny2016}. This behavior is expected for the leading order drag effect \cite{Levchenko2008,Schuett2013}, which is equivalent to the kinetic theory approach used to obtain the transconductivity \cite{Narozhny2012}. In particular, in the strong screening limit
$\kappa\ll1$ the drag resistivity in the Fermi-liquid regime depends
on the carrier density as $\rho_{D}\propto1/\left(n_{a}n_{p}\right)^{3/2}$ \cite{Tse2007}, regardless of the temperature difference between the layers.

\section{Results and discussion}\label{sec:results}

In general, the integration of Eq. (\ref{eq:resistivity}) can only be done numerically. We will consider relatively large interlayer separations ($dk_{F}\gg1$), so that any heat transfer between the layers is
kept at a minimum. For typical carrier densities $n\sim10^{12}\;\mathrm{cm^{-2}}$,
the condition $dk_{F}\gg1$ is satisfied for $d\gtrsim50$ nm. Furthermore, since the dielectric constant of aerogel materials is very low \cite{Hrubesh1993,Hrubesh1998}, we consider a coupling constant $\alpha\sim2$ in graphene, close to its vacuum value $\sim2.2$ \cite{CastroNeto2009}. We note, however, that many-body effects within each graphene layer may renormalize $\alpha$ towards lower values \cite{DasSarma2007,Borghi2009}.

\subsection{Numerical results}\label{subsec:numerical}

The numerical results for the drag resistivity are shown in Fig. \ref{fig:2}. The density plot in Fig. \ref{fig:2}(a) illustrates that the drag depends  significantly on the temperature difference between the graphene layers. It also clearly shows that the drag behavior is not symmetrical when the layer temperatures are interchanged. In particular, when $T_a\gg T_p$ the drag resistivity is larger than when $T_a\ll T_p$. 

When the temperature $T_j$ of one layer is kept fixed as the temperature $T_i$ of the other layer is varied, the drag resistivity as a function of $T_i$ clearly deviates from the usual quadratic temperature dependence $\rho_D\propto T_i^2$ [Fig. \ref{fig:2}(b)]. The effect is more pronounced when $T_i\ll T_j$, which implies low temperatures $T_i$. This behavior should be contrasted with the known
crossover to a linear dependence $\rho_{D}\propto T$ in the usual
case $T_{a}=T_{p}$, which only occurs at relatively high temperatures \cite{Lux2012,Chen2015,Escudero2022}. Note that Fig. \ref{fig:2}(b) is not symmetrical under a temperature exchange, in the sense that varying $T_p$ while keeping $T_a$ constant yields a different drag resistivity magnitude.

Figure \ref{fig:2}(c) shows the drag resistivity when both $T_{a}$ and $T_{p}$ vary, but in such a way that there is always a constant temperature difference $\Delta T=T_{p}-T_{a}$. This situation may represent temperature fluctuations in a conventional experiment, where the ideal situation $T_{a}=T_{p}$ may not be exactly
realized. The resulting deviations from $\rho_{D}\propto T^{2}$ can still be
quite significant, even at relatively small temperature differences between
the layers. In general, considering $\rho_D$ as a function of $T_{a}$, the deviations
from the dependence $\rho_{D}\propto T_{a}^{2}$ are more pronounced
when $\left|\Delta T\right|$ is comparable to (or larger than) $T_{a}$,
which, in practical terms, implies low temperatures $T_{a}$.

Significant deviations from the behavior $\rho_D\propto T^2$ in the Fermi-liquid regime are usually attributed to the appearance of other drag mechanisms, such as the frictional drag mediated by phonon-electron interactions \cite{Boensager1998,Amorim2012a,Fandan2019}. In contrast, the deviations from the $T^2$ behavior shown in Fig. \ref{fig:2} arise only from the effect of a temperature difference between the layers on the drag relaxation rate. In graphene, the phonon-mediated drag becomes important only at relatively high temperatures $T\gtrsim150$ K \cite{Amorim2012a}. For the Coulomb drag at equal layer temperatures, the deviation to a linear dependence $\rho_D\propto T$ begins to take place at roughly $\lambda=dk_{B}T/\hbar v_{F}\sim0.2$ \cite{Lux2012}, although it becomes significant only for $\lambda\gtrsim1$, which for $d=100$ nm implies $T\gtrsim 75$ K [see Fig. \ref{fig:2}(b)]. Consequently, within the Fermi-liquid regime, significant deviations from the $T^2$ behavior at low temperatures can only occur when the graphene layers are at different temperatures.

Although we have focused on the Coulomb drag between two graphene monolayers, our results are also relevant for the Coulomb drag between two, two-dimensional electron gases \cite{Gramila1991,Zheng1993}, two graphene bilayers  \cite{Hwang2011,Lux2013,Li2016}, or different layered materials \cite{Scharf2012,Zhu2020}. Equation (\ref{eq:transconductivity}) is readily generalized to these cases, provided that one uses the corresponding dispersion relation, form factor, and scattering time. Since Eq. (\ref{eq:NLsus}) is in general temperature dependent (even within the degenerate limit), the quantitative effect of a temperature difference between the layers on the drag resistivity is expected to be system dependent. The universal quadratic temperature behavior $\rho_D\propto T^2$ (i.e., regardless of the electronic properties of the layers) holds only at equal layer temperatures, where the transconductivity only depends on the nonlinear susceptibility (\ref{eq:NLsus0}), which to leading order is temperature independent in the Fermi-liquid regime. In this sense, the Coulomb drag depends more strongly on the electronic properties of the layers if they are at different temperatures. 

\subsection{Analytical approximations}\label{subsec:analytical} 

The temperature dependence of the drag when $T_{a}\ne T_{p}$ can
be understood by analyzing the limiting behaviors of the function
(\ref{eq:G}), for the limits $\gamma\ll1$ and $\gamma\gg1$. Since we are considering $dk_F\gg1$ and $\alpha\sim2$, the factor $\kappa\ll1$ in Eq. (\ref{eq:resistivity}) will be ignored \cite{Lux2012,Escudero2022}.

In the Fermi-liquid regime, where $k_{B}T_{a}/\mu\ll1$, the case $\gamma\ll1$
necessarily implies low temperatures in the active layer, so that
$\lambda\ll1$ (technically, if $dk_{F}$ is very large, one may have
$\lambda\sim1$ even if $k_{B}T_{a}/\mu$ is very small; however, this
implies very large interlayer separations and carrier densities, a
special situation which we shall neglect). Similarly, the opposite
case $\gamma\gg1$ implies high values of $\lambda$. Thus for $\gamma\ll1$
the integrand in Eq. (\ref{eq:G}) can be expanded to lowest order in $\gamma$,
whereas for $\gamma\gg1$ the integrand can be expanded to lowest
order in $y$. [The latter approximation holds because the $x$-integral
in Eq. (\ref{eq:resistivity}) implies a cutoff at low values of $x$.] We then have,
to leading order in $x$, 
\begin{equation}
	G\left(x,\lambda,\gamma\right)\simeq\frac{4x}{3\lambda}\begin{cases}
		1 & \gamma\ll1,\\
		2\gamma & \gamma\gg1.
	\end{cases}
\end{equation}
Replacing in Eq. (\ref{eq:resistivity}) yields
\begin{equation}
	\rho_{D}\simeq-\frac{h}{e^{2}}\frac{\pi^{3}}{720}\frac{k_{B}/\mu}{\left(dq_{T}\right)^{2}\left(dk_{F}\right)^{3}}\begin{cases}
		T_{p}/2 & \gamma\ll1,\\
		T_{a} & \gamma\gg1.
	\end{cases}\label{eq:dragaprox}
\end{equation}
Thus, when $T_{i}\gg T_{j}$, the phase space available for electron-hole
pairs is limited by the higher temperature $T_{i}$. Both limits
of $\gamma$ also imply a stronger dependence on the interlayer separation,
$\rho_{D}\propto1/d^{5}$, which is the same behavior obtained when
$T_{a}=T_{p}$, at large temperatures. Note that the drag resistivity
in the limit $T_{a}\gg T_{p}$ is twice as large as in the opposite
limit [see Fig. \ref{fig:2}(a)].

\section{Conclusions}\label{sec:conclusions}

We have theoretically studied the Coulomb drag between two graphene layers at different temperatures. Our analysis holds for doped graphene layers with equal carrier density, within the degenerate limit $\mu/k_BT\gg1$ and in the ballistic regime. In terms of the temperatures $T_a$ and $T_p$ of the active and passive layers, respectively, we have found significant deviations from the usual quadratic temperature dependence in the drag resistivity when $T_a\neq T_p$. In particular, in the limit $T_i\gg T_j$ the drag resistivity depends only on the much higher temperature as $\rho_D\propto T_i$. The magnitude of this behavior is not symmetrical: when $T_a\gg T_p$ the drag resistivity is about twice as large as in the opposite case. This behavior holds in general, in the sense that $\rho_D$ is not symmetrical when the temperatures of the layers are interchanged. Furthermore, the limit $T_i\gg T_j$ also implies a stronger dependence $\rho_D\propto 1/d^5$ on the interlayer separation $d$. The obtained results can be pronounced even at relatively small temperature differences between the layers. Thus, experimental deviations from the usual behavior $\rho_D\propto T^2$ may be attributed to thermal fluctuations. 

\begin{acknowledgments}
This paper was partially supported by grants of CONICET (Argentina National Research Council) and Universidad Nacional del Sur (UNS) and by ANPCyT through PICT 2019-03491 Res. No. 015/2021, and PIP-CONICET 2021-2023 Grant No. 11220200100941CO. J.S.A. acknowledges support as a member of CONICET, F.E. acknowledges support from a research
fellowship from this institution, and F.A. acknowledges support as a member of Departamento de Física, Universidad
Nacional del Sur.
\end{acknowledgments}

\onecolumngrid
\appendix

\section{Transconductivity between two graphene layers at different temperatures}\label{appendix:transconductivity}

The current can be separated by its contributions from intraband impurity
scattering and interlayer scattering \cite{Hwang2011}, $\mathbf{j}_{i}=\mathbf{j}_{i}^{\mathrm{imp}}+\mathbf{j}_{i}^{\mathrm{inter}}$,
where $\mathbf{j}_{i}^{\mathrm{imp}}=\sigma_{ii}\mathbf{E}_{i}$ and
$\mathbf{j}_{i}^{\mathrm{inter}}=ge\sum_{\mathbf{k},s}\mathbf{v}_{\mathbf{k},s}\tau_{\mathbf{k}}^{i}\mathcal{I}_{\mathbf{k},s}^{i,\mathrm{inter}}$.
Here $\mathcal{I}_{\mathbf{k},s}^{i,\mathrm{inter}}$ is the collision
integral due to the interlayer interaction:
\begin{align}
	\mathcal{I}_{\mathbf{k},s}^{i,\mathrm{inter}} & =\sum_{\mathbf{k}',\mathbf{q}}\sum_{s',r,r'}\mathcal{W}_{s,s',r,r'}\left(\mathbf{k},\mathbf{k}',\mathbf{q}\right)\left[\tilde{f}_{\mathbf{k}+\mathbf{q},s'}^{i}\tilde{f}_{\mathbf{k}'-\mathbf{q},r'}^{j}\left(1-\tilde{f}_{\mathbf{k},s}^{i}\right)\left(1-\tilde{f}_{\mathbf{k}',r}^{j}\right)\right.\nonumber \\
	& \qquad\left.-\tilde{f}_{\mathbf{k},s}^{i}\tilde{f}_{\mathbf{k}',r}^{j}\left(1-\tilde{f}_{\mathbf{k}+\mathbf{q},s'}^{i}\right)\left(1-\tilde{f}_{\mathbf{k}'-\mathbf{q},r'}^{j}\right)\right]\delta\left(\epsilon_{\mathbf{k},s}+\epsilon_{\mathbf{k}',r}-\epsilon_{\mathbf{k}+\mathbf{q},s'}-\epsilon_{\mathbf{k}'-\mathbf{q},r'}\right),
\end{align}
where $\mathcal{W}_{s,s',r,r'}\left(\mathbf{k},\mathbf{k}',\mathbf{q}\right)$
is the interlayer scattering probability for $\left|\mathbf{k},s;\mathbf{k}',r\right\rangle \rightarrow\left|\mathbf{k}+\mathbf{q},s';\mathbf{k}'-\mathbf{q},r'\right\rangle $.
Within Fermi's golden rule,
\begin{equation}
	\mathcal{W}_{s,s',r,r'}\left(\mathbf{k},\mathbf{k}',\mathbf{q}\right)=\frac{2\pi g}{\hbar}\left|U_{ap}\left(\mathbf{q}\right)\right|^{2}F_{s,s'}\left(\mathbf{k},\mathbf{q}\right)F_{r,r'}\left(\mathbf{k}',-\mathbf{q}\right).
\end{equation}
Replacing Eq. (\ref{eq:noneq}), using conservation of energy, and the relation
\begin{align}
	f_{\mathbf{k}+\mathbf{q},s'}^{i}f_{\mathbf{k}'-\mathbf{q},r'}^{j}\left(1-f_{\mathbf{k},s}^{i}\right)\left(1-f_{\mathbf{k}',r}^{j}\right) & =f_{\mathbf{k},s}^{i}f_{\mathbf{k}',r}^{j}\left(1-f_{\mathbf{k}+\mathbf{q},s'}^{i}\right)\left(1-f_{\mathbf{k}'-\mathbf{q},r'}^{j}\right)e^{-\Delta\beta\Delta\epsilon},\label{eq:App1}
\end{align}
where $\Delta\beta=\beta_{i}-\beta_{j}$ and $\Delta\epsilon=\epsilon_{\mathbf{k}+\mathbf{q},s'}-\epsilon_{\mathbf{k},s}$,
gives $\mathbf{j}_{i}^{\mathrm{inter}}=\mathbf{j}_{i,0}^{\mathrm{inter}}+\mathbf{j}_{i,1}^{\mathrm{inter}}$,
where (to linear order in $\psi$) 
\begin{align}
	\mathbf{j}_{i,0}^{\mathrm{inter}} & =-ge\sum_{\mathbf{k},\mathbf{k}',\mathbf{q}}\sum_{s,s',r,r'}\mathcal{W}_{s,s',r,r'}\left(\mathbf{k},\mathbf{k}',\mathbf{q}\right)\mathbf{v}_{\mathbf{k},s}\tau_{\mathbf{k}}^{i}f_{\mathbf{k},s}^{i}f_{\mathbf{k}',r}^{j}\left(1-f_{\mathbf{k}+\mathbf{q},s'}^{i}\right)\left(1-f_{\mathbf{k}'-\mathbf{q},r'}^{j}\right)\nonumber \\
	& \qquad\times\left(e^{-\Delta\beta\Delta\epsilon}-1\right)\delta\left(\epsilon_{\mathbf{k},s}+\epsilon_{\mathbf{k}',r}-\epsilon_{\mathbf{k}+\mathbf{q},s'}-\epsilon_{\mathbf{k}'-\mathbf{q},r'}\right),\\
	\mathbf{j}_{i,1}^{\mathrm{inter}} & =-ge\sum_{\mathbf{k},\mathbf{k}',\mathbf{q}}\sum_{s,s',r,r'}\mathcal{W}_{s,s',r,r'}\left(\mathbf{k},\mathbf{k}',\mathbf{q}\right)\mathbf{v}_{\mathbf{k},s}\tau_{\mathbf{k}}^{i}f_{\mathbf{k},s}^{i}f_{\mathbf{k}',r}^{j}\left(1-f_{\mathbf{k}+\mathbf{q},s'}^{i}\right)\left(1-f_{\mathbf{k}'-\mathbf{q},r'}^{j}\right)\nonumber \\
	& \qquad\times\left\{ \left(\psi_{\mathbf{k},s}^{i}+\psi_{\mathbf{k}',r}^{j}-\psi_{\mathbf{k}+\mathbf{q},s'}^{i}-\psi_{\mathbf{k}'-\mathbf{q},r'}^{j}\right)+\left(e^{-\Delta\beta\Delta\epsilon}-1\right)\left[f_{\mathbf{k},s}^{i}\psi_{\mathbf{k},s}^{i}+f_{\mathbf{k}',r}^{j}\psi_{\mathbf{k}',r}^{j}\right.\right.\nonumber \\
	& \qquad\left.\left.-\psi_{\mathbf{k}+\mathbf{q},s'}^{i}\left(1-f_{\mathbf{k}+\mathbf{q},s'}^{i}\right)-\psi_{\mathbf{k}'-\mathbf{q},r'}^{j}\left(1-f_{\mathbf{k}'-\mathbf{q},r'}^{j}\right)\right]\right\} \delta\left(\epsilon_{\mathbf{k},s}+\epsilon_{\mathbf{k}',r}-\epsilon_{\mathbf{k}+\mathbf{q},s'}-\epsilon_{\mathbf{k}'-\mathbf{q},r'}\right).
\end{align}
The contribution $\mathbf{j}_{i,0}^{\mathrm{inter}}$, which is independent
of the electric fields (does not depend on $\psi$) and only appears
when $\Delta\beta\neq0$, vanishes when integrated over $\mathbf{q}$.
Then, replacing $\psi_{\mathbf{k},s}^{i}=\beta_i e\mathbf{E}_{i}\cdot\tau_{\mathbf{k}}^{i}\mathbf{v}_{\mathbf{k},s}$,
using the relations
\begin{align}
	f_{\mathbf{k},s}^{i}f_{\mathbf{k}',r}^{j}\left(1-f_{\mathbf{k}+\mathbf{q},s'}^{i}\right)\left(1-f_{\mathbf{k}'-\mathbf{q},r'}^{j}\right) & =-\frac{e^{\Delta\beta\Delta\epsilon/2}\left(f_{\mathbf{k},s}^{i}-f_{\mathbf{k}+\mathbf{q},s'}^{i}\right)\left(f_{\mathbf{k}',r}^{j}-f_{\mathbf{k}'-\mathbf{q},r'}^{j}\right)}{4\sinh\left(\beta_{i}\Delta\epsilon/2\right)\sinh\left(\beta_{j}\Delta\epsilon/2\right)},\\
	\mathcal{W}_{s,s',r,r'}\left(\mathbf{k},\mathbf{k}',\mathbf{q}\right) & =\mathcal{W}_{s',s,r',r}\left(\mathbf{k}+\mathbf{q},\mathbf{k}'-\mathbf{q},-\mathbf{q}\right),
\end{align}
and standard manipulations \cite{Jauho1993}, yields the transconductivity given in the main text.

\section{Calculation of $\tilde{\boldsymbol{\Gamma}}_{i}\left(\mathbf{q},\omega\right)$}\label{appendix:NLS}

Considering a momentum-dependent scattering time $\tau_{\mathbf{k}}^{i}=\tau_{i}k$,
and only the conduction band, Eq. (\ref{eq:NLsus}) reads
\begin{align}
	\tilde{\boldsymbol{\Gamma}}_{i}\left(\mathbf{q},\omega\right) & =2\pi g\tau_{i}v_{F}\mathbf{q}\sum_{\mathbf{k}}\left(f_{\mathbf{k}}^{i}-f_{\mathbf{k}+\mathbf{q}}^{i}\right)f_{\mathbf{k}+\mathbf{q}}^{i}F\left(\mathbf{k},\mathbf{q}\right)\delta\left(\hbar\omega+\epsilon_{\mathbf{k}}-\epsilon_{\mathbf{k}+\mathbf{q}}\right)\nonumber \\
	& +2\pi g\tau_{i}v_{F}\sum_{\mathbf{k}}\mathbf{k}\left(f_{\mathbf{k}}^{i}-f_{\mathbf{k}+\mathbf{q}}^{i}\right)\left(f_{\mathbf{k}}^{i}+f_{\mathbf{k}+\mathbf{q}}^{i}-1\right)F\left(\mathbf{k},\mathbf{q}\right)\delta\left(\hbar\omega+\epsilon_{\mathbf{k}}-\epsilon_{\mathbf{k}+\mathbf{q}}\right).
\end{align}
The calculation will be done for $\omega>0$ . Taking into account
the change of angle $\theta=\theta_{\mathbf{k}}-\theta_{\mathbf{q}}$,
we have
\begin{equation}
	\mathbf{k}\rightarrow\frac{\mathbf{q}}{q}k\cos\theta+\frac{\mathbf{e}_{z}\times\mathbf{q}}{q}k\sin\theta.
\end{equation}
Due to the delta functions, the term proportional to $\sin\theta$
vanishes after the integration over $\theta$.
Then, writing 
\begin{equation}
	F\left(\mathbf{k},\mathbf{q}\right)=\frac{1}{2}\left(1+\frac{k+q\cos\theta}{\left|\mathbf{k+q}\right|}\right),
\end{equation}
the integration of the delta over the angle $\theta$ gives
\begin{equation}
	\tilde{\boldsymbol{\Gamma}}_{i}\left(\mathbf{q},\omega\right)=\theta\left(v_{F}q-\omega\right)\frac{g\tau_{i}v_{F}}{2\pi\hbar v_{F}}\frac{\mathbf{q}}{\sqrt{\left(q^{2}-\tilde{\omega}^{2}\right)}}\left[X_{i}\left(\mathbf{q},\omega\right)+Y_{i}\left(\mathbf{q},\omega\right)\right],
\end{equation}
where
\begin{align}
	X_{i}\left(\mathbf{q},\omega\right) & =\int_{\left(q-\tilde{\omega}\right)/2}^{\infty}dk\left(f_{k}^{i}-f_{k+\tilde{\omega}}^{i}\right)f_{k+\tilde{\omega}}^{i}\sqrt{\left(\tilde{\omega}+2k\right)^{2}-q^{2}},\\
	Y_{i}\left(\mathbf{q},\omega\right) & =\int_{\left(q-\tilde{\omega}\right)/2}^{\infty}dk\left(f_{k}^{i}-f_{k+\tilde{\omega}}^{i}\right)\left(f_{k}^{i}+f_{k+\tilde{\omega}}^{i}-1\right)\sqrt{\left(\tilde{\omega}+2k\right)^{2}-q^{2}}\frac{\tilde{\omega}^{2}+2k\tilde{\omega}-q^{2}}{2q^{2}},
\end{align}
with $\tilde{\omega}=\omega/v_{F}$. Replacing the relations (\ref{eq:aprox1}) and (\ref{eq:aprox2}) yields
\begin{align}
	X_{i}\left(\mathbf{q},\omega\right) & =\theta\left(k_{F,i}-\frac{q+\tilde{\omega}}{2}\right)\frac{k_{F,i}}{\beta_{i}\mu_{i}}\tanh\left(\frac{\beta_{i}\hbar\omega}{2}\right)\sqrt{\left(\tilde{\omega}-2k_{F,i}\right)^{2}-q^{2}},\\
	Y_{i}\left(\mathbf{q},\omega\right) & =\sum_{\ell=\pm}\theta\left(k_{F,i}-\frac{q+\ell\tilde{\omega}}{2}\right)\ell\frac{k_{F,i}}{\beta_{i}\mu_{i}}\sqrt{\left(\tilde{\omega}-\ell2k_{F,i}\right)^{2}-q^{2}}\frac{-\ell\tilde{\omega}^{2}+2k_{F,i}\tilde{\omega}-q^{2}}{2q^{2}}.
\end{align}
Expanding the above expressions to lowest order in $\omega$ and $q$
results in Eq. (\ref{eq:NLsus2}).

\twocolumngrid

\end{document}